\documentclass[preprint,pre,aps,showpacs]{revtex4}
\usepackage{graphicx}
\begin{document}

\title{Effect of secondary decay on isoscaling: Results from the canonical
thermodynamical model} 

\author{Gargi Chaudhuri and Swagata Mallik}

\affiliation{Variable Energy Cyclotron Centre, 1/AF Bidhannagar, Kolkata
700064}

\date{\today}

\begin{abstract}
 The projectile fragmentation reactions using $^{58}Ni$ $\&$ $^{64}Ni$ beams at
 140 MeV/n on targets
 $^{9}Be$ $\&$ $^{181}Ta$ are studied using the
 canonical thermodynamical model coupled with an evaporation code.
 The isoscaling property of the fragments produced is studied using both the primary and
 the secondary fragments and it is observed that the secondary fragments also
 respect isoscaling though the isoscaling parameters $\alpha$ and $\beta$ changes. The
 temperature needed to reproduce experimental data with the secondary fragments
 is less than that needed with the primary ones. The canonical model coupled
 with the evaporation code successfully explains the experimental data for
 isoscaling for the
 projectile fragmentation reactions.

\end{abstract}


\maketitle

\section{Introduction}

Projectile fragmentation reaction is used extensively to study the reaction
mechanisms in heavy ion collisions at intermediate and high energies. This is
also  an important technique  for the production of rare isotope beams and is
used by many radioactive ion beam facilities around the world.  The  fragment
cross sections of projectile fragmentation reactions using primary beams of
$^{40}Ca$, $^{48}Ca$, $^{58}Ni$ and $^{64}Ni$  at 140 MeV/nucleon on $^{9}Be$
 and $^{181}Ta$ targets have been
measured at the National Superconducting Cyclotron Laboratory at Michigan State
University\cite{Mocko1}. The canonical thermodynamical model(CTM)\cite{Das1} has been used to
calculate some of these fragment cross sections\cite{Gargi1}.  In the present
work, an evaporation code has been
developed and has been coupled with the canonical thermodynamical model. CTM
coupled with this secondary decay code is  then used to analyze the isoscaling data from the
projectile fragmentation reactions.  The lighter fragments produced
from these reactions exhibit the linear isoscaling\cite{Xu,Tsang1,Tsang2} phenomena and our model calculation also strongly supports
this observation. The secondary fragments (produced after applying the
evaporation code on the canonical model) also exhibit isoscaling like  the 
primary fragments from the fragmentation reaction\cite{Gargi2} but the
temperature needed to reproduce the  experimental data with the secondary
fragments  is lower than that 
required by the calculation with the primary fragments. The isoscaling parameters $\alpha$
 and $\beta$ as obtained in the present work from the model calculations agree closely with those obtained from
the experimental data. These parameters as obtained from the secondary fragments
are lower in magnitude than those obtained from the primary ones. This effect is
also seen in the dynamical models \cite{Ono1} though the reduction is much
more there. The  isoscaling behaviour displayed by the fragments produced in the
projectile fragmentation reactions  and the effect of evaporation on it in the
framework of the HIPSE model has been discussed recently in \cite{Fu}.

This paper is structured as follows. First we describe the canonical model
briefly  in Sec.II. In the same section we also present the main features of
the evaporation code which is used to calculate the secondary fragments. In 
Sec.III we present the results.  The effect of sequential decay on the distribution
of the isotopic fragments is discussed. The isoscaling
phenomena as displayed by the primary as well as the secondary fragments is 
also
described in this section and are compared with the experimental data. In Sec.
IV we present the summary.

\section{The Statistical Model}
In models of statistical disassembly of a nuclear system formed by 
the collision of two heavy ions at intermediate energy one assumes
that because of multiple nucleon-nucleon collisions a statistical
equilibrium is reached. Consequently, the temperature rises.  The system expands
from normal density and composites are formed on the way to disassembly.
As the system reaches between three to six times the normal volume, the
interactions between composites become unimportant (except for the
long range Coulomb interaction) and one can do a statistical equilibrium
calculation to obtain the yields of composites at a volume called the
freeze-out volume.  The partitioning into available channels can be
solved in the canonical ensemble where the number of particles in the
nuclear system is finite (as it would be in experiments). In the next subsection
we  describe the canonical model.

\subsection{The canonical thermodynamical model}
 In this section we describe briefly the canonical thermodynamical model.    Assume that the system with 
 $A_0$ nucleons and $Z_0$
protons at temperature $T$, has expanded to a higher than normal volume 
and the partitioning into different composites can be calculated according
to the rules of equilibrium statistical mechanics.  In a canonical model, the partitioning
is done such that all partitions have the correct $A_0, Z_0$ (equivalently
$N_0, Z_0$).  Details of the implementation of the canonical model
can be found elsewhere \cite{Das1}; here we give the essentials
necessary to follow the present work.

The canonical partition function is given by
\begin{eqnarray}
Q_{N_0,Z_0}=\sum\prod \frac{\omega_{I,J}^{n_{I,J}}}{n_{I,J}!}
\end{eqnarray}
Here the sum is over all possible channels of break-up (the number of such
channels is enormous) which satisfy $N_0=\sum I\times n_{I,J}$
and $Z_0=\sum J\times n_{I,J}$; $\omega_{I,J}$ 
is the partition function of one composite with
neutron number $I$ and proton number $J$ respectively and $n_{I,J}$ is
the number of this composite in the given channel.  
The one-body partition
function $\omega_{I,J}$ is a product of two parts: one arising from
the translational motion of the composite and another from the
intrinsic partition function of the composite:
\begin{eqnarray}
\omega_{I,J}=\frac{V_f}{h^3}(2\pi mT)^{3/2}A^{3/2}\times z_{I,J}(int)
\end{eqnarray}
Here $A=I+J$ is the mass number of the composite and
$V_f$ is the volume available for translational motion; $V_f$ will
be less than $V$, the volume to which the system has expanded at
break up. We use $V_f = V - V_0$ , where $V_0$ is the normal volume of  
nucleus with $Z_0$ protons and $N_0$ neutrons.  In this calculation we
have used a fairly typical value $V=6V_0$.

The probability of a given channel $P(\vec n_{I,J})\equiv P(n_{0,1},
n_{1,0},n_{1,1}......n_{I,J}.......)$ is given by
\begin{eqnarray}
P(\vec n_{I,J})=\frac{1}{Q_{N_0,Z_0}}\prod\frac{\omega_{I,J}^{n_{I,J}}}
{n_{I,J}!}
\end{eqnarray}
The average number of composites with $I$ neutrons and $J$ protons is
seen easily from the above equation to be
\begin{eqnarray}
\langle n_{I,J}\rangle=\omega_{I,J}\frac{Q_{N_0-I,Z_0-J}}{Q_{N_0,Z_0}}
\end{eqnarray}
The constraints $N_0=\sum I\times n_{I,J}$ and $Z_0=\sum J\times n_{I,J}$
can be used to obtain different looking but equivalent recursion relations
for partition functions\cite{Chase}.  For example
\begin{eqnarray}
Q_{N_0,Z_0}=\frac{1}{N_0}\sum_{I,J}I\omega_{I,J}Q_{N_0-I,Z_0-J}
\end{eqnarray}
These recursion relations allow one to calculate $Q_{N_0,Z_0}$ 

We list now the properties of the composites used in this work.  The
proton and the neutron are fundamental building blocks 
thus $z_{1,0}(int)=z_{0,1}(int)=2$ 
where 2 takes care of the spin degeneracy.  For
deuteron, triton, $^3$He and $^4$He we use $z_{I,J}(int)=(2s_{I,J}+1)\exp(-
\beta E_{I,J}(gr))$ where $\beta=1/T, E_{I,J}(gr)$ is the ground state energy
of the composite and $(2s_{I,J}+1)$ is the experimental spin degeneracy
of the ground state.  Excited states for these very low mass
nuclei are not included.  
For mass number $A=5$ and greater we use
the liquid-drop formula.  For nuclei in isolation, this reads ($A=I+J$)
\begin{eqnarray}
z_{I,J}(int)=\exp\frac{1}{T}[W_0A-\sigma(T)A^{2/3}-\kappa\frac{J^2}{A^{1/3}}
-C_s\frac{(I-J)^2}{A}+\frac{T^2A}{\epsilon_0}]
\end{eqnarray}
The derivation of this equation is given in several places
\cite{Bondorf1,Das1}
so we will not repeat the arguments here.  The expression includes the 
volume energy, the temperature dependent surface energy, the Coulomb
energy and the symmetry energy.  The term $\frac{T^2A}{\epsilon_0}$
represents contribution from excited states
since the composites are at a non-zero temperature.

We also have to state which nuclei are included in computing $Q_{N_0,Z_0}$ 
(eq.(17)).
For $I,J$, (the neutron and the proton number)
we include a ridge along the line of stability.  The liquid-drop
formula above also gives neutron and proton drip lines and 
the results shown here include all nuclei within the boundaries.

The long range Coulomb interaction between
different composites can be included in an approximation called
the Wigner-Seitz approximation.  We incorporate this following the
scheme set up in \cite{Bondorf1}.  

\subsection{The evaporation code}
The statistical multifragmentation model described above calculates the
properties of the collision averaged system that can be approximated by an
equilibrium ensemble. Ideally, one would like to measure the properties of
excited primary fragments after emission in order to extract information about
the collisions and compare directly with the equilibrium predictions of the
model. However, the time scale of a nuclear reaction($10^{-20}s$) is much
shorter than the time scale for particle detection ($10^{-9}s$). Before reaching
the detectors, most fragments decay to stable isotopes in their ground states.
Thus before any model simulations can be compared to experimental data, it is
indispensable to have a model that simulates sequential decays.
A Monte Carlo technique is employed to follow all decay chains until the
resulting products are unable to undergo further decay. For the purposes of the
sequential decay calculations  the excited primary fragments generated by the
statistical model calculations are taken as the compound nucleus input to the
evaporation code. Hence, every primary fragment is decayed as a separate event.

We consider the deexcitation of a primary fragment of mass $A$, charge $Z$
and temperature $T$. The succseesive particle emission from  the hot primary fragments
is assumed to be the basic deexcitation mechanism. For each event of the primary
breakup simulation, the entire chain of evaporation and secondary breakup events
is Monte Carlo simulated. The standard Weisskopf evaporation
 scheme is used to take into account evaporation of nucleons, $d$, $t$,
 $He^3$ and $\alpha$. The decays of particle stable excited states via gamma
 rays were also taken into account for the sequential decay process and for the
 calculation of the final ground state yields. We have also considered fission
 as a deexcitation channel though for the nuclei of mass $<$ 100 its role will be
quite insignificant. The process of light particle emission from a compound nucleus is
governed by the emission width $\Gamma_{\nu}$ at which a
particle of type $\nu$ is emitted. According to
Weisskopf's conventional evaporation theory \cite{Weiss}, the
partial decay width for emission of a light particle of type $\nu$
 is given by
 \begin{equation}
 \Gamma_{\nu} = \frac {gm\sigma_0}{\pi^{2}\hbar^{2}} \frac {(E^*-E_0-V_\nu)}
 {a_R} \exp({2 \sqrt{a_R(E^*-E_0-V_\nu)}-2\sqrt{a_PE^*}})
 \end{equation}

 Here  $m$ is the mass of the emitted particle, $g$ is its spin degeneracy. 
 $E_0$ is the particle separation energy which is calculated from the binding
 energies of the parent nucleus, daughter nucleus and the binding energy of the
 emitted particle and the liquid drop model is used to calculate the binding
 energies. The subscript $\nu$ refers to the emitted particle, $P$
 refers to the parent nuclei and $R$ refers to the residual(daughter) nuclei. 
 $a_P$  $\&$ $a_R$ are the level
 density parameters of the parent and residual nucleus respectively. The level
 density  parameter is given by $a = A/16  MeV^{-1}$ and it connects the
 excitation energy $E^*$ and temperature $T$ through the following relations.

 \begin{eqnarray}
 E^* = a_PT_P^2 \nonumber\\
 (E^*-E_0-V_\nu) = a_RT_R^2.
 \end{eqnarray}

 where $T_P$ $\&$ $T_R$ are the temperatures of the emitting(parent) and the
 final(residual) nucleus respectively. $V_\nu$ is the Coulomb barrier which is
 zero for neutral particles and non-zero for charged particles. In order to calculate the
 Coulomb barrier for charged particles of mass $A \ge 2$ we use a touching sphere
 approximation\cite{Friedman},
 
 \begin{eqnarray}
 V_\nu &=&
 \frac{Z_\nu(Z_P-Z_\nu)e^2}{r_i\{A_\nu^{1/3}+{(A_P-A_\nu)}^{1/3}\}}\hspace{1cm}
 \mbox{for} \hspace{1cm} A_\nu \ge 2\nonumber\\
  &=&  \frac{(Z_P-1)e^2}{r_iA_P^{1/3}}  \hspace{2.9cm} \mbox{for}
 \hspace{1cm} protons
 \end{eqnarray}
 
 where $r_i$ is taken as 1.44m. 
 
 $\sigma_0$ is the geometrical crosssection (inverse cross section) associated with the formation of the
 compound nucleus(parent) from the emitted particle and the daughter nucleus and
 is given by $\sigma_0=\pi R^2$ where,

 \begin{eqnarray}
  R &=& r_0\{{(A_P-A_\nu)}^{1/3} + {A_\nu}^{1/3}\}
 \hspace{1.2cm}\mbox{for}\hspace{0.3cm} A_\nu \ge 2\nonumber\\
    &=& r_0({A_P-1})^{1/3} \hspace{2.8cm}\mbox{for}\hspace{0.3cm} A_\nu = 1.
  \end{eqnarray}

where $r_0$ = 1.2 fm.

For the emission of giant dipole $\gamma$-quanta we take the
formula given by Lynn\cite{Lynn}
\begin{equation}
\Gamma_{\gamma}={3 \over \rho_{P}(E^{*})}\int_{0}^{E^*}d\varepsilon\rho_{R}(E^*-\varepsilon)f(\varepsilon)\label{4l}
\end{equation}
with
\begin{equation}
f(\varepsilon)= {4 \over 3\pi}{1+\kappa \over m_{n}c^{2}}{e^{2} \over
\hbar c}{N_PZ_P \over A_P} {\Gamma_{G}\varepsilon^{4} \over
(\Gamma_{G}\varepsilon)^{2}+(\varepsilon^{2}-E_{G}^{2})^{2}}\label{4m}
\end{equation}
with $\kappa=0.75$, and $E_{G}$ and $\Gamma_{G}$ are the position
and width of the giant dipole resonance. 

For the fission  width we have used the simplified formula of
Bohr-Wheeler given by 

\begin{equation}
\Gamma_f = \frac{T_P}{2\pi}\exp{(-B_f/T_P)}
\end{equation}

where $B_f$ is the fission barrier of the compound nucleus given
by\cite{Guaraldo}

\begin{equation}
 B_f (MeV)= -1.40Z_P + 0.22(A_P-Z_P) +101.5.
 \end{equation}
 
 Once the emission widths are known, it is required to
establish the emission algorithm which decides  whether a particle is
 being emitted from the compound nucleus.
 This is done \cite{thesis} by first
calculating the ratio $x=\tau / \tau_{tot}$  where $\tau_{tot}=
\hbar  / \Gamma_{tot}$, $\Gamma_{tot}=\sum_{\nu}\Gamma_{\nu}$ and
$\nu = n,p,d,t,He^3,\alpha,\gamma$ or fission and then performing Monte-Carlo
sampling from a uniformly distributed set of random numbers. 
 In
the case that a particle is emitted, the type of the emitted
particle is next decided by a Monte Carlo selection with the
weights $\Gamma_{\nu}/\Gamma_{tot}$ (partial widths). 
 The energy of the emitted particle is then obtained by
another Monte Carlo sampling of its energy spectrum. The energy, mass and charge
of the nucleus is adjusted after each emission. This procedure is followed for
each of the primary fragment produced at a fixed temperature and then
repeated over a large ensemble and the observables are calculated from the
ensemble averages. . The number and type of particles emitted and the
final decay product in each event is registered and are taken into account  properly
keeping in mind the overall charge and baryon number conservation.

\section{Results}
First we will show our calculations for $^{58}Ni$ on $^{9}Be$ reaction and 
$^{64}Ni$ on $^{9}Be$ reaction. In the model, the target imparts a certain amount of
energy to the projectile transforming it to a projectile like fragment(PLF) with
a temperature. This excited PLF will then expand and form composites during the
expansion. The partioning of the PLF into different composites is done by the
rules of equilibrium statistical mechanics in a freeze-out volume.  We consider production of different isotopes from
the statistical breakup of the dissociating system. If $<n_{i,j}>$ is the  average
number(multiplicity) of composites with $i$ neutrons and $j$ protons, then the
cross-section for this composite is 
$\sigma(i,j)=C<n_{i,j}>$, where C is a constant not calculable from the
thermodynamic model.  It depends upon the dynamics that are outside the scope of
this model.  To be able to compute $<n_{i,j}>$ we need to know the mass and
charge of the PLF and its temperature. The source sizes adopted for this calculation are zero order
guesses. It could be sometimes smaller or greater depending on the diffusion
from the target.  For $^{64}Ni$ or $^{58}Ni$ on $^{9}Be$ which is  a small
target the choice of the mass and the charge of the PLF is limited. It can be
slightly less than that of the projectile to as large as that of the projectile
plus $^{9}Be$, the last being the case when the much larger projectile swallows the
small target and drags it along retaining PLF features. Similarly we have some
limits on energy imparted(this fixes the temperature). This energy can be small
or upto the upper limit. The upper limit is given by the case of projectile
swallowing Be and all the energy transforming into internal excitation(no part
going into collective flow). In the canonical calculation, the dissociating
system is taken to be $^{58}Ni+^{9}Be(N_0=35,Z_0=32)$ and for the other reaction the
dissociating system is taken to be $^{64}Ni+^{9}Be(N_0=41,Z_0=32)$.All composites between
drip lines are included as detailed in Sec.IIA with the highest values of
N, Z terminating at $N_0$, $Z_0$. The temperature is taken to be 5.8 MeV for both the
reactions. 

Fig. 1 displays the isotopic distribution for Z=12(magnesium) produced from both
the the
reactions. The dashed lines correspond to the distributions of the primary
fragments while the solid lines correspond to the distributions after sequential
decay. As expected, the more neutron rich system with $N_0/Z_0=1.28 $ (right
panel) produces
more neutron rich isotopes than the neutron deficient system with $N_0/Z_0=1.09$
(left panel).
In all cases, the primary distributions are much wider and more neutron rich
than the final distributions.  The peak positions of the isotopic distributions of
both the primary and the secondary fragments coincide in case of the neutron
deficient system as seen from the left panel of the figure. In case of the
neutron rich system(right panel) the peak of the distribution of the secondary
fragments has shifted to the left with respect to that of the primary.
 The experimental isotopic distributions(solid squares with error bars) agree much more
with the final results obtained after secondary decay than with the primary
distributions. The width and peak position of the isotopic distribution after
the secondary decay agrees very well with the experimental data. The model also
successfully reproduces the rapid fall in crosssection for large neutron number.

We will now discuss the results about isoscaling. It is observed from the
experimental data\cite{Mocko2}  that the light
fragments  emitted from the $^{58}Ni$ and $^{64}Ni$ systems exhibit the linear 
isoscaling behaviour represented by the equation
\begin{equation}
R_{21}= Y_2(N,Z)/Y_1(N,Z)=C\exp(\alpha N+\beta Z).
\end{equation}
where the isoscaling ratio $R_{21}(N,Z)$ is factored into two fugacity terms
$\alpha$ and $\beta$, which contain the differences of the chemical potentials
for neutrons and protons of the two reaction systems. $Y_{2}(N,Z)$ refers to the
yield of fragment(N,Z) from system 2 which is usually taken to be the
neutron-rich one and $Y_{1}(N,Z)$ refers to the same from system 1.  C is a normalization factor of
the isoscaling ratio.  It is observed from our model that both the primary as well as the secondary fragments  
 exhibit isoscaling. Fig. 2 shows the isoscaling results for Ni on Be system for the
 primary  fragments. The ratio $R_{21}$ is plotted as function of the neutron
 number from Z =6 to Z= 13 in the left panel whereas the right panel displays
 the ratio as function of the proton number Z from N=8 to N=15. It is seen that 
 the primary fragments exhibit very well the linear
 isoscaling behaviour for the lighter fragments  over a wide range of
 isotopes and isotones. The lines in the figures are the best fits of the calculated $R_{21}$
 ratios(open triangles) to Eq.15. They are essentially linear and parallel on 
 the semi log plot .

 Fig. 3  displays the isoscaling results for the secondary fragments. The open
 triangles are the results obatined from our model while the solid squares with the
 error bars are the experimental ratios. We have shown the isoscaling
results for the even Z and odd Z isotopes in two separate panels for the sake of
clarity.
 While comparing with the results of the primary
fragments in Fig 2, it is evident that  the isoscaling is valid for a limited range of
isotopes for the secondary fragments as compared to the primary ones.  When the isotopes away
from the valley of stable nuclei  are considered, the trends for the secondary
fragments are not as clearly
consistent with the isoscaling law as are the trends of the primary
distribution.  One can conclude from this that isoscaling is approximately valid
in the case of the secondary fragments. The lines 
in the figures are the best fits of
the calculated $R_{21}$ ratios which agree closely
with the experimental data.                         
The temperature required to reproduce isoscaling data with the primary fragments
is about 8 MeV\cite{Gargi2} whereas that required for the secondary fragments is 5.8 MeV. 
This decrease was already predicted in one of our earlier papers\cite{Gargi2} in
Sec.9.  It has also been found out by Ono et al.\cite{Ono1} that the effect of secondary
decay is to decrease the isoscaling parameter $\alpha$ by about $50\%$. This is indeed what
emerges from our calculations after including secondary decay code with the
canonical model though the amount of reduction is less as compared to the
dynamical model.  Tha value of $\alpha$ as obtained from the fits of the primary
fragments (left panel of Fig. 2) is 0.713 whereas that obtained from the 
secondary fragments is 0.580
which is much closer to  the
experimentally obtained value for $\alpha$\cite{Mocko2} equal to 0.566.  

 In Fig. 4 we have also plotted the isoscaling ratios for different 
neutron number
values as function of  the proton number Z and thereby calculated the other isoscaling parameter
$\beta$ from them. The value of $\beta$ as obatined from the linear fits of the
primary fragments( right panel of Fig. 2) is -0.849. For the secondary fragments,
 the value of $\beta$ from our
model is -0.634 whereas the experimentally obtained value is -0.621. The value
of $\beta$ also decreases in the case of secondary from the primary ones as in the
case of $\alpha$.  The results
obtained after the sequential decay matches very well with the experimental
ratios.  As for the isotopes in Fig 3,  for the isotones also it is seen
that the linear isoscaling is valid for a limited range in case of secondary
fragments as compared to the primary ones(right panel of Fig.2).

We now turn to the case where the target is $^{181}Ta$. One can consider the 
 extreme limit which is target
independence and by which we mean that $N_0$, $Z_0$ refers to simply to the case
where just the projectile is the disintegrating system. The more likely scenario
where some matter has diffused to or from the target has many possibilities. In
principle, the target could shear away some material from the projectile leaving
a PLF which is a fraction of the projectile. For a peripheral collision this is
less likely than the alternative of the projectile picking up part of nuclear
matter from the tail region of the much larger target. The amount of nuclear
matter curved out of Ta will be small(for the disintegrating system to retain
PLF characteristics) but other than that not much can be said and an integration
over the different possibilities might be essential. Keeping this limitation in
mind, we compared with different possible scenarios and the case with projectile
plus 10 neutrons and 8 protons from the target, i.e, projectile +$^{18}0$ yielded the
best results. It is also seen that target independence gives worse results than
this case.  The agreement of our model
calculations with experimental data is pretty good in this case. The temperature used for
this reaction is 6.2 MeV.  
 
Fig. 5  and Fig. 6 are similar to Fig. 3 \& Fig. 4 except for the fact that they
are for the case where the target is $^{181}Ta$ instead of $^{9}Be$. The
theoretical slope $\alpha$ as obtained from Fig. 5 from the fit of the 
secondary fragments is 0.459
while the experimentally obtained slope is 0.432. The straight line fit to the
calculated points matches nicely with the experimental ratios. The value of
$\beta$ as obtained from our model from the slopes in Fig. 6 is -0.489 whereas the
experimentally obtained value is -0.487. Thus we find that the values of the
isoscaling parametrs as obtained from the secondary fragments for both the
targets agree quite well with the experimental values as can be seen from Table
1.

\begin{table}[!htbp]
  \begin{center}
    \begin{tabular}{c|c|c|c|c|c|}
    \hline
       & Target material & Isoscaling parameters &  primary & secondary & experiment \\
      \hline
        & $^{9}Be$ & $\alpha$ ($Z_{min}=6,Z_{max}=13$) & 0.713 & 0.580 & 0.566  \\
	\hline
	& $^{9}Be$ & $\beta$ ($N_{min}=8,N_{max}=15$) & -0.849 & -0.634 & -0.621  \\
	\hline
       & $^{181}Ta$ & $\alpha$($Z_{min}=6,Z_{max}=13$) & 0.619 & 0.459 & 0.432 \\
       \hline
	& $^{181}Ta$ & $\beta$ $N_{min}=8,N_{max}=15$) & -0.682 & -0.489 &
	-0.487  \\
      \hline
    \end{tabular}
  \end{center}
  \caption{ Best fit values of the isoscaling parametrs $\alpha$ and $\beta$ for
  the two targets $^{9}Be$ and $^{181}Ta$. The values obtained from the slope of
  the primary and secondary fragments as well as the experimental values are
  tabulated. In the second column the range of $Z$ or $N$ values used to
  calculate the parameters are indicated.}
 
  \label{tab1}
\end{table}
\section{Summary}
This work deals with the developing of the sequential decay
code and successfully coupling it with the canonical thermodynamical model in
order to compare the properties of the secondary fragments with the
experimental data. The width, peak position and rapid fall in cross-section of
the isotopic distribution of the secondary fragments matches well with the
experimental data. The main purpose is to examine the effects of sequential decay on the
phenomenon of isoscaling. The secondary fragments also shows isoscsling and the 
the isoscaling parameters calculated  from the secondary fragments matches 
closely with the experimental data. The temperature required to reproduce the
experimental data with the secondary fragments is less than needed by the
primary ones.   We finally 
conclude that the
canonical thermodynamical model
 can explain the isoscaling properties of the
lighter fragments produced in the projectile fragmentation reaction.

\section{Acknowledgement}
The authors gratefully acknowledges important discussions with Prof. Subal
Dasgupta. They are thankful to   Prof. M.B. Tsang and Dr. M. Mocko. Valuable
suggestions from Dr. Santanu Pal and Jhilam Sadhukhan is also acknowledged
gratefully.

\begin{figure}[htbp]
\includegraphics[width=6.0in,height=4.5in,clip]{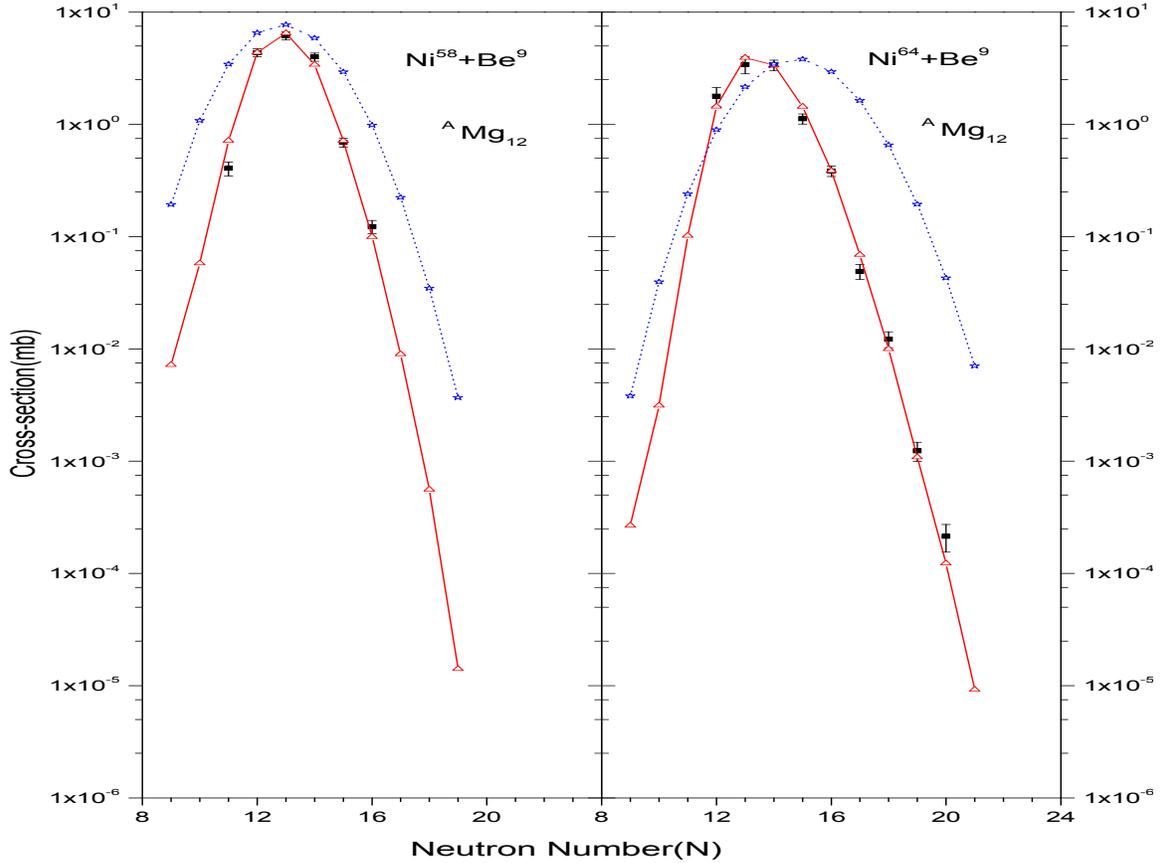}
\caption{ Experimental cross sections of magnesium isotopes(squares with error
bars) compared with theoretical results: primary  fragments( open stars joined by
dotted line) and secondary fragments(open triangles joined by solid line). The left
panel is for the reaction $^{58}$Ni on $^{9}$Be while the right panel is for
$^{64}$Ni on $^{9}$Be reaction. The temperature is 5.8 MeV for both the
reactions.}
\label{fig1}
\end{figure}
\newpage

\begin{figure}
\includegraphics[width=6.0in,height=4.5in,clip]{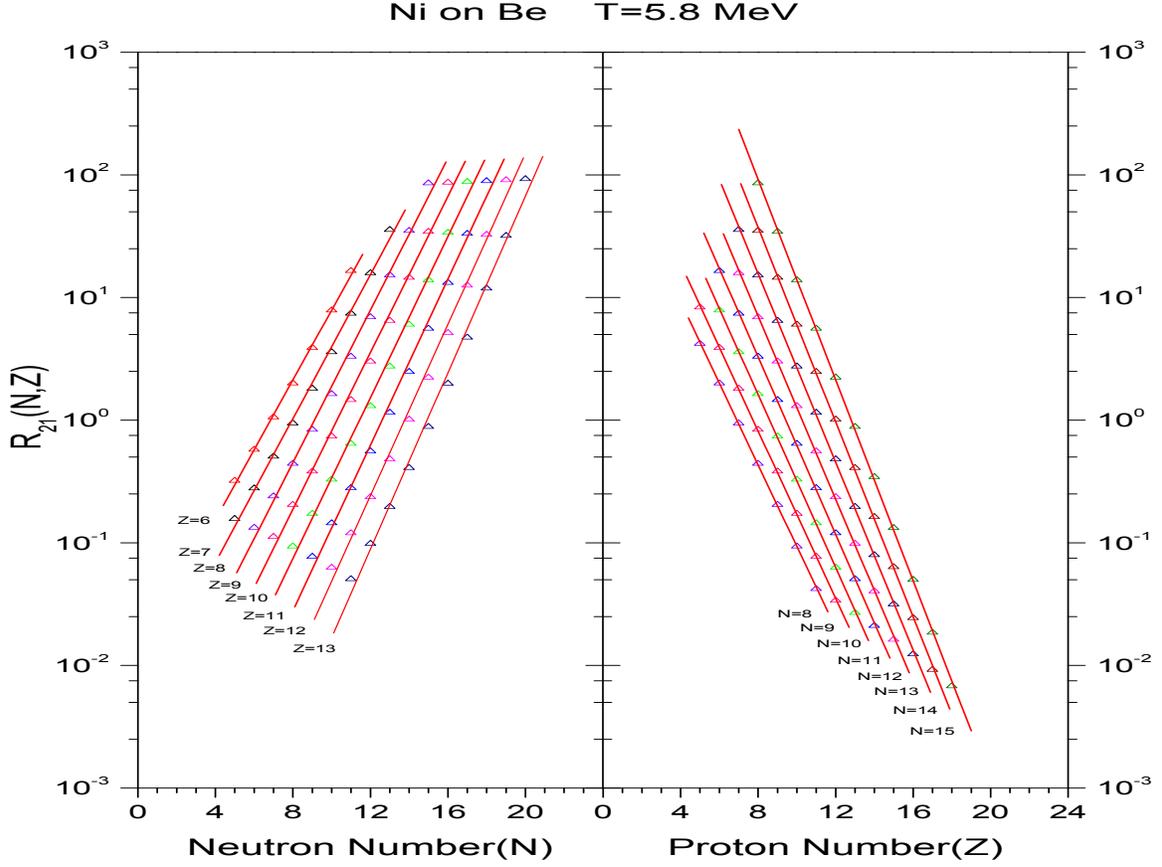}
\caption{ Ratios($R_{21}$) of multiplicities of primary fragments of producing the nucleus
$(N,Z)$ where reaction 1 is  $^{58}$Ni on $^{9}$Be and reaction 2 is $^{64}$Ni 
on $^{9}$Be. The left panel shows the ratios as function of neutron number $N$
for fixed $Z$ values from 6 to 13, while the right panel displays the ratios as
function of proton number $Z$ for fixed neutron numbers from $N$ = 8 to 15. The lines drawn through the theoretical
points(open triangles) are best fits of the calculated ratios. The
temperature used for both the reacions is 5.8 MeV.} 
\label{fig2}
\end{figure}

\begin{figure}[htbp]
\includegraphics[width=6.0in,height=4.5in,clip]{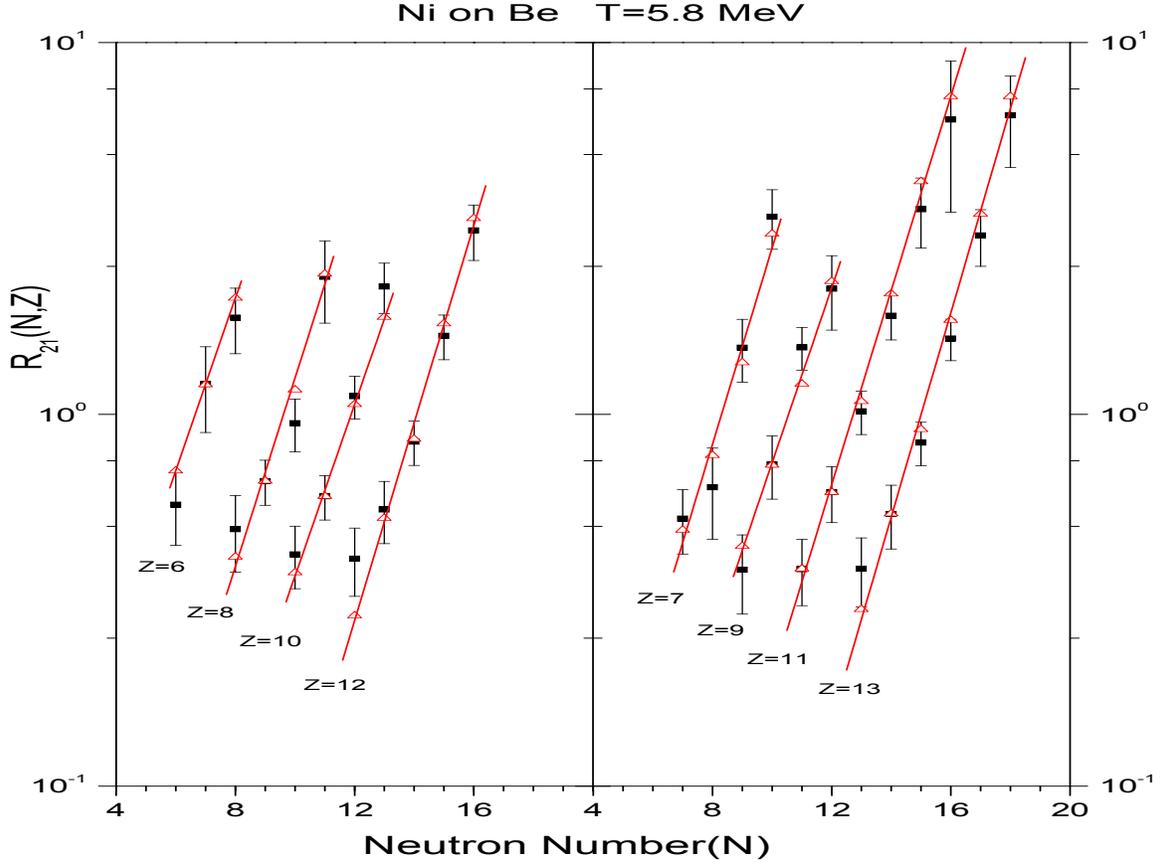}
\caption{ Ratio of multiplicities of  the secondary fragments of producing the nucleus
$(N,Z)$ where reaction 1 is  $^{58}$Ni on $^{9}$Be and reaction 2 is $^{64}$Ni 
on $^{9}$Be compred with the ratios of the experimental cross sections of the
same two reactions. The left panel shows the even $Z$ isotopes while the right
panel shows the results for the odd ones. The lines drawn through the theoretical
points(open triangles) are best fits of the calculated ratios. The experimental
points are shown by solid squares with error bars.The
temperature used for both the reacions is 5.8 MeV. } 
\label{fig3}
\end{figure}

\begin{figure}[htbp]
\includegraphics[width=6.0in,height=4.5in,clip]{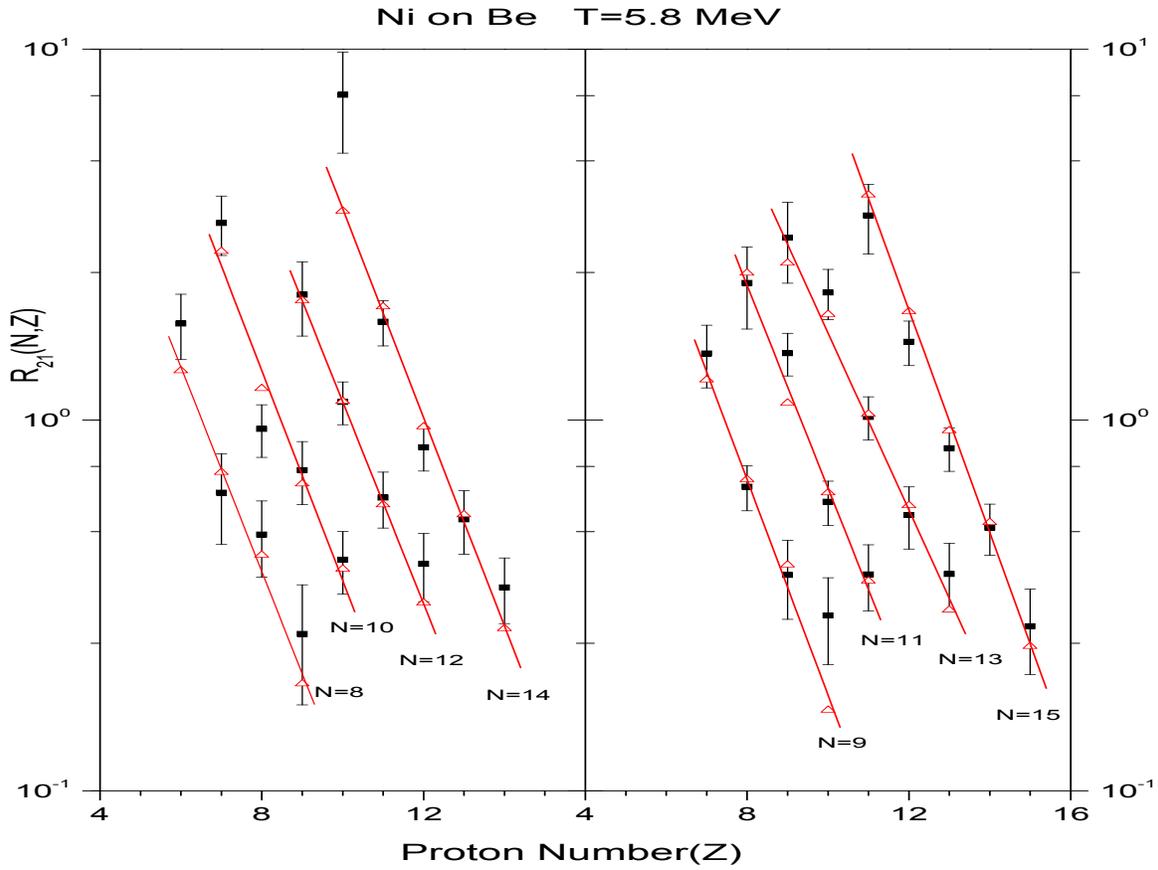}
\caption{ Same as Fig. 3 except that the ratios are plotted as function of the
proton number $Z$ for fixed neutron numbers. The left panel shows the results for
the even neutron numbers while the right ones show those for the odd ones. } 
\label{fig4}
\end{figure}

\begin{figure}[htbp]
\includegraphics[width=6.0in,height=4.5in,clip]{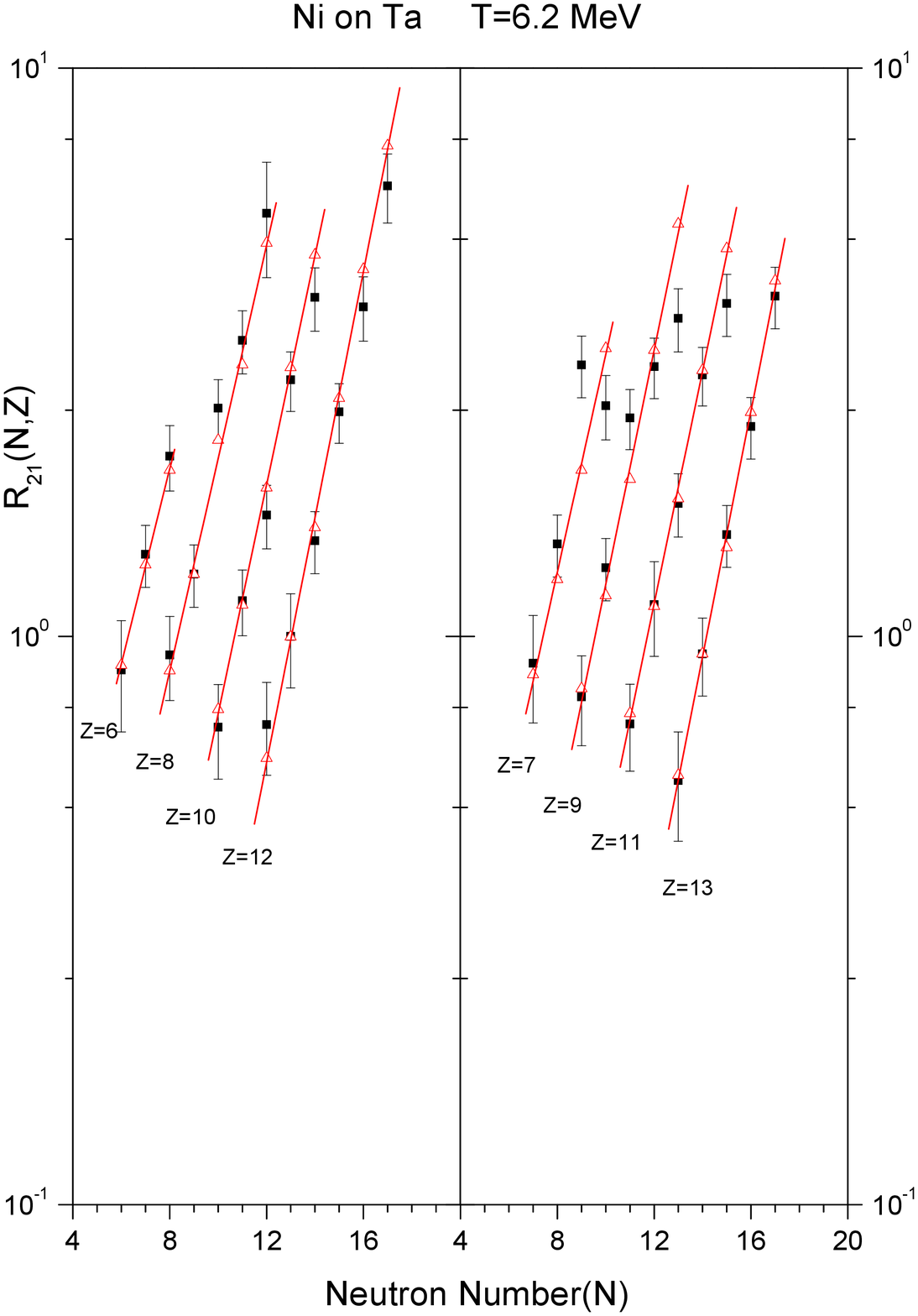}
\caption{ Same as Fig. 3 except that here the target is $^{181}Ta$ instead of
$^{9}Be$. The temperature used for both the the reactions is 6.2 MeV. } 
\label{fig5}
\end{figure}

\begin{figure}[htbp]
\includegraphics[width=6.0in,height=4.5in,clip]{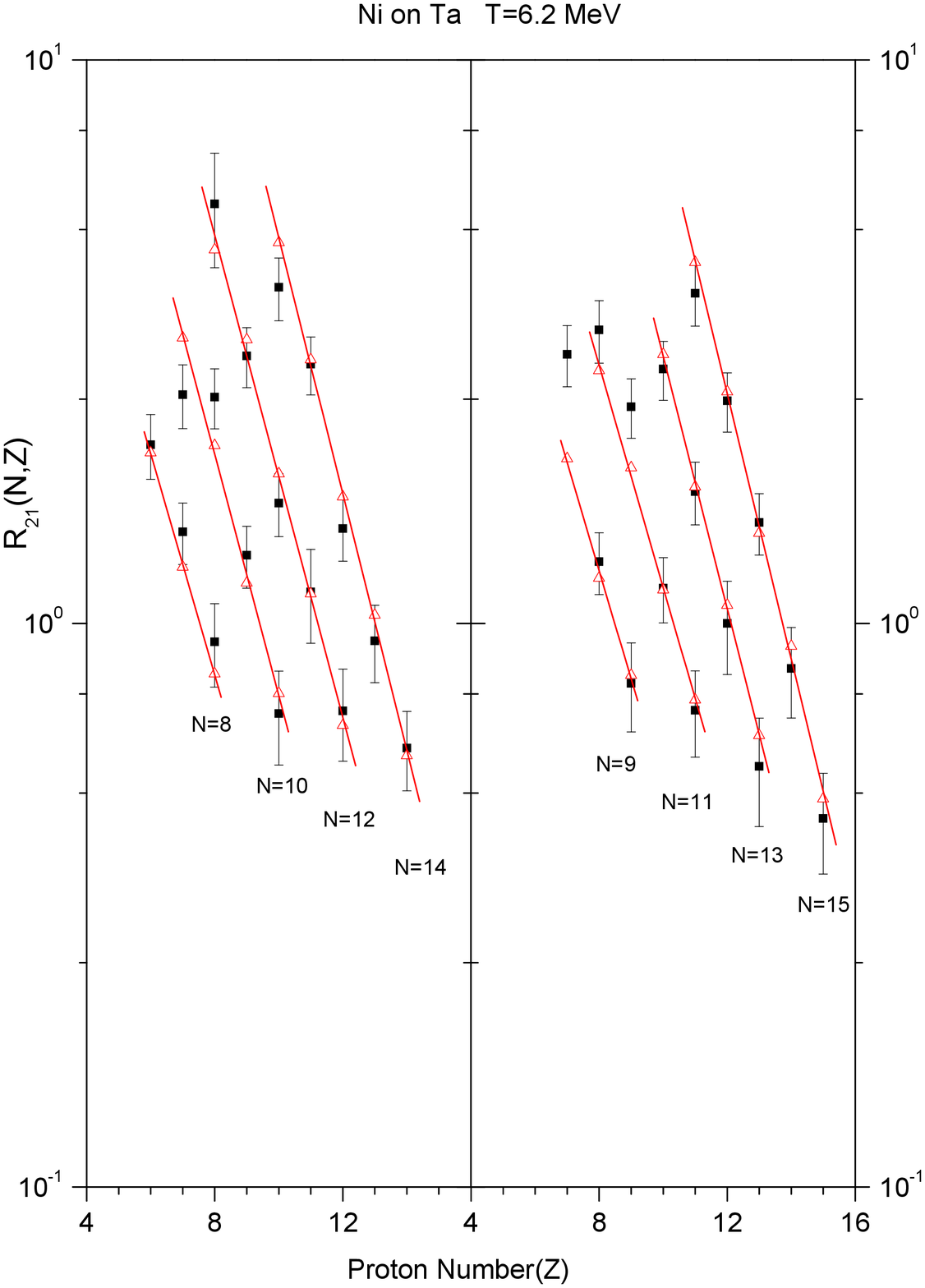}
\caption{ Same as Fig. 4 except that here the target is $^{181}Ta$ instead of
$^{9}Be$. The temperature used for both the reactions is 6.2 MeV. } 
\label{fig6}
\end{figure}

\end{document}